# Deterministic synthesis of phase pure 2D perovskites via progressive transformation of layer thickness

Jin Hou[1], Wenbin Li[2,3], Hao Zhang[2,3], Siraj Sidhik[1], Jean-Christophe Blancon[2], Jacky Even[4], Mercouri G. Kanatzidis[5] and Aditya D. Mohite[1,2*]

**[1]Department of Materials Science and NanoEngineering, Rice University, Houston, Texas 77005, USA.**

**[2]Department of Chemical and Biomolecular Engineering, Rice University, Houston, Texas 77005, USA.**

**[3]Applied Physics Graduate Program, Smalley-Curl Institute, Rice University, Houston, TX, 77005, USA.**

**[4]Fonctions Optiques pour les Technologies de l'Information (FOTON), INSA de Rennes, CNRS, UMR 6082, 35708 Rennes, France.**

**[5]Department of Chemistry and Department of Materials Science and Engineering, Northwestern University, Evanston, Illinois 60208, USA.**

**Two-dimensional (2D) halide perovskites have emerged as semiconductor platforms for realizing efficient and durable optoelectronic devices. However, the reproducible synthesis of 2D perovskite crystals with desired layer thickness (or n value) greater than 2, has been an enduring challenge due to the lack of kinetic control (temperature, time, stoichiometry) for each layer thickness. Here, we demonstrate a novel method term as the kinetically controlled space confinement (KCSC) for the deterministic growth of phase pure Ruddlesden-Popper (RP) and Dion-Jacobson (DJ) 2D perovskites. The phase-pure growth was achieved by progressively increasing the temperature (fixed time) or the crystallization time (fixed temperature), which allowed for an acute control of the crystallization kinetics. We also observe a systematic transformation from a lower n-value to n=3, 4, 5, 6 in 2D perovskites. In-situ photoluminescence spectroscopy and imaging suggest that the progressive transformation from lower to higher n-value occurs via intercalation of excess precursor ions. These experiments enabled the development of a machine learning assisted multi-parameter phase diagram, which predicts the growth of 2D phase with a specific n-value.**

Two-dimensional halide perovskites (2D-HaP) are a sub-class of three-dimensional HaP, which have emerged as a new class of highly durable solution-processed organic-inorganic (hybrid) low-dimensional semiconductors.[1–3] They exhibit a unique combination of properties, which are derived from four exciting classes of materials - quantum wells,[4] atomically thin 2D materials[5], organic semiconductors[6,7], and 3D-HaP perovskites[8,9]. The general chemical formula of 2D-HaP is $(A')_m(A)_{n-1}M_nX_{3n+1}$ (where A′ is a bulky organic cation, A is a small organic cation, M is a divalent metal, X is a halide, m=2 in RP phase and m=1 in DJ phase, n determine the thickness of the inorganic layer), which consists of alternate layers of organics $(A')_m$ and inorganic $(A)_{n-1}M_nX_{3n+1}$, providing a perfect platform to study organic-inorganic interfaces. There is growing consensus that their physical properties are dictated by the interaction between the organic cation and the inorganic framework, which presents a unique opportunity to tailor their behaviors. For examples, charge-carrier mobility,[10,11] nonlinear optical effects,[12] electron-phonon coupling,[13] ferroelectricity,[14,15] and Rashba effect[16], etc.

However, most of these studies on 2D HaP have been performed on lower n-values (n=1, 2). We suggest there is difficulty in reproducibly growing higher n-value phase pure crystals (n>3).[5,12,17–19], thus resulting in mixed phase single crystals and powders observed as multiple peaks in the absorption spectrum. We hypothesize that the impurity phases (multiple n-values) largely arise from the lack of control of crystallization kinetics (temperature, time) using the classical method commonly used for the growth of 2D HaP crystals and powders. Briefly, the classical method involves dissolving all precursors in a solvent at elevated temperature (230 °C) to achieve supersaturation followed by rapid crystallization through fast-cooling.[1–3,20]Once the temperature is lowered to a point where supersaturation diminishes, crystal nucleation dominates crystal growth, resulting in large amount of small crystals in a very short period of time (half hour). Therefore, decoupling crystallization temperature and time is difficult in the classical synthesis method since there is no freedom to control the temperature during the crystallization process. Typically, in classic synthesis the crystallization time or rate is not monitored because it assumes those kinetics parameters do not impact phase purity. However, their impact is largely underestimated, a systematic study on the effect on decoupled temperature and time of crystallization is still missing.

Moreover, the powder-form of 2D-HaP is limited by its small crystal size (μm to mm sizes) making it challenging to perform experiments such as, (i) Light-matter interactions; [21–23] (ii) Optical measurement and dielectric functions where large-area flat crystal is required;[4] (iii) Electron-phonon coupling, where carrier trapping and exciton dissociation at grain boundaries can often be complicated spectra.;[24] There has been several efforts to grow large area 2D HaP crystals by increasing the crystallization temperature and further dilution, but with little success in fabricating higher n-values (n≥4) which are more desirable due to their smaller band gaps, and better charge transport properties.[1,20,25] The previous strategies to obtain high n-value crystal involves, (i) isolation of n-pure crystallite form a n-mixed crystal[20] or (ii) large number of crystallization attempts. However, a direct and reproducible approach to synthesize phase-pure high n-value crystals is needed.

Here, we demonstrate the deterministic growth of phase-pure Ruddlesden-Popper (RP) and Dion-Jacobson (DJ) 2D perovskite crystals achieved via a systematic transformation from a lower n-value to n=3, 4, 5, 6. We show that both the solution crystallization time and temperature, besides the stoichiometry of precursors, significantly impacts the n-value of 2D-HaP crystal. By increasing the time of crystallization and/or the temperature of synthesis, the resulting 2D-HaP crystal transforms from lower n-values to higher n-values, indicated by in-situ absorbance and x-ray diffraction measurements. To understand the underlying mechanism, we developed a novel approach termed as kinetically controlled space confinement (KCSC) method, which enables the precise control over the annealing time and temperature, which in turn slows down the crystallization rate thus allowing for precise control over the synthesis of the desired n-value.[26,27] In-situ photoluminescence spectroscopy and imaging results demonstrate the progressive transformation from lower to higher n-value, which we propose is due to the intercalation of excess precursor ions at the organic-inorganic interface of 2D-HaP. Based on the experimental data using the KCSC method, we created a multi-parameter phase diagrams for both the RP and DJ 2D-HaP, predicting the growth of each n-value at specific time and temperature using machine learning (ML), which is applicable to all 2D-HaP structural phases. Finally, we show that the transformation effect can be translated to the classical synthesis used commonly across the research community to synthesize powders of 2D-HaP. We believe that a direct, reproducible approach to acquire phase-pure single crystal 2D-HaP for all n-values (n=1-6) would enable a wide-spread investigation of 2D HaPs and their incorporation beyond passivation layers or as additives for bulk perovskite films. Importantly, as previously demonstrated, the use of phase pure 2D HaPs into photovoltaic devices and other optoelectronic devices will enable highly durable devices based on 2D HaPs and their incorporation into 3D HaPs[28,29].

**Result and discussion:**

The progressive transformation is demonstrated via the KCSC method, which is shown in Fig. 1a. Compared to classic synthesis, the concentration of solution used in KCSC method is diluted two times to avoid supersaturation, allowing for a slow, gradual growth of one seed instead of a rapid, massive nucleation, hence the time window of crystallization is prolonged. The diluted solution is sandwiched between two pre-heated glass substrates which is placed on the hotplate for annealing at different temperatures. Millimeter to centimeter-sized single crystal (~500nm thick) will be acquired (Fig. S1-S3) after several hours. This KCSC method enables us to fine-tune the kinetics parameters (crystallization temperature and crystallization time). Interestingly, we observe that further annealing on the crystal in the parent solution will make it evolve to higher n-value crystal.

We performed in-situ optical and X-ray spectroscopy measurements on the RP 2D-HaP crystal fabricated by the KCSC method. Figure.1b illustrates the evolution of optical absorption as a function of temperature (ranging from 60-100 ◦C) with a fixed time of 7 hours. To keep all other variables constant such as the stoichiometry of the precursors, we used the same parent solution for each KCSC growth. Results indicate a strong correlation between temperature and the *n* value of the RP 2D-HaP crystal. At low temperatures (60-66 ◦C), the KCSC method produces pure *n*=3 crystals which further goes through a mixture of n=3, and n=4 at 68-79 ◦C, following which pure n=4 2D HaP forms. At high temperatures (80-90 ◦C), the KCSC method produced pure *n*=4 and then mixture of n=4 and n=5 and finally n=5 when the temperature approaches 100 ◦C. The in-situ x-ray diffraction measurements (Fig. 1d) validate the same trend. Next, we altered the crystallization time by keeping the temperature fixed at 75 °C and tracked the evolution of a single perovskite crystal growth after each time interval. Figure 1c shows the absorbance spectra as a function of time for the KCSC method. Results indicated a similar transformation trend as the temperature dependence data with an intermediate mixed phase between the *n*=3 and *n*=4 perovskite on its way to forming a pure n=5. The time dependent absorption results imply that there is a full transformation of the crystal from a lower *n* value to higher *n* value. This is further supported by XRD measurement as shown in Fig. 1d, 1e, and Fig. S4).

In order to understand the underlying transformation/intercalation process of the 2D-HaP single crystal during synthesis, we perform an in-situ photoluminescence (PL) imaging during the growth

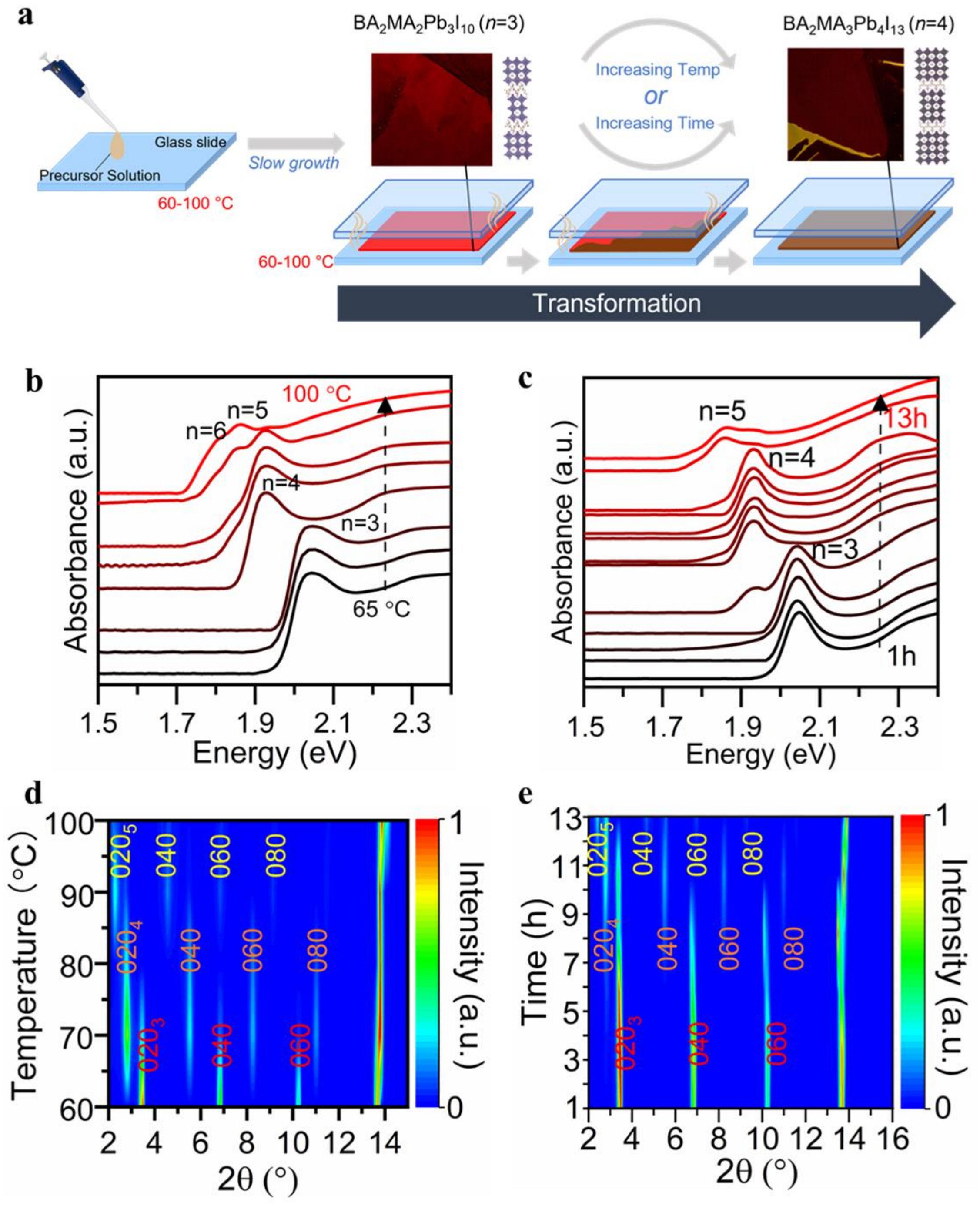

**Fig. 1 | Transformation of 2D RPPs (*n*=3) in KCSC.** (a) Schematic illustration of transformation of RP n=3 to n=4 in KCSC method, diluted parent n=3 solution was put between two substrates and upon heating, n=3 crystal formed first, then it transformed to n=3 and n=4 mixed crystal, finally pure n=4. Temperature dependent in-situ absorbance (b) and (d) 1D X-Ray spectrum showed the n increases as the temperature increases. Time dependent in-situ absorbance (c) and 1D X-Ray spectrum(e) with evolution of XRD spectral intensity shoed *n* increase as time increases.

process (Fig.2). The PL setup is shown in fig. 2a, where we measure the emission of the perovskite crystal by exciting the whole region using a mercury vapor lamp (U-LH100HG, Olympus) with a 532 nm short pass filter mount before the sample. The PL was collected by adding a Semrock Razoredge 633nm long pass filter right before the CCD camera. The filter suppresses the emission from the RP n=3 perovskite and allows for the detection of the n=4 phase. Figure 2c illustrates the evolution of the 2D-HaP single crystal as a function of time at 75 °C. Specifically, we monitor the edge of the crystal during growth since intercalation in 2D-HaP first occurs at this position. We propose that small precursor ions, such as $MA^+$, $Pb^{2+}$, and $I^-$ penetrate the lattice from the edges of the 2D-HaP crystal and diffuse along the interface between the inorganic perovskite layers. These ions fill the voids of the corner sharing $PbI_6$ structure, forming an additional linkage of the $(MA)PbI_6$ lattice. This templating effect occurs because of the weak van der Waals interaction between the 2D inorganic octahedra sheets. In order to understand the transformation process and its underlying mechanisms, we probed the photoluminescence spectra as a function of the growth time as shown in Fig. 2b. The circle illustrated on the PL imaging data (fig. 2c) is the probe region with a laser beam diameter of 30 μm under a 10x objective. The spectra show a progressive change in emission peak from **630nm** to **650nm** indicating an increase in layer thickness from n=3 to n=4 which agrees with our hypothesis. In addition, the photoluminescence spectra show no other emission sites, which excludes the possibility of defects and phase impurity. Lastly, both PL imaging and PL spectra results indicate that the transformation occurs much faster on the edges than in the bulk. We also probed the PL spectra from the back side of the crystal and the identical *n* was observed (fig.S5), which excluded the possibility that transformation is from high *n*-value crystal forming around lower n-value to form vertically stacked heterostructures. Figure 2d illustrates the proposed intercalation growth mechanism for the RP-HaP. Here, the solution first forms the n=3 perovskite which is shown by the stacking of 3 inorganic $[PbI_6]^{4-}$ layers separated by the butylammonium organic spacer cation. Then over time, the organic MA, Pb and I molecule diffuse through the edges of the n=3 2D-HaP crystal and attaches to the $PbI_6$ structure as the terminal group. The intercalation process results in the increase of the layer thickness from n=3 to n=4 for the synthesized 2D-HaP. Our results are consistent with previous works that observe the intercalation of precursor ions into the lattice to form 2D-HaP and higher layer thickness 2D-HaP [30,31]. We also considered a related inverse mechanism where large cations (in this case BA) escaping from the perovskites structure, is excluded since it lead to the collapse of the structure

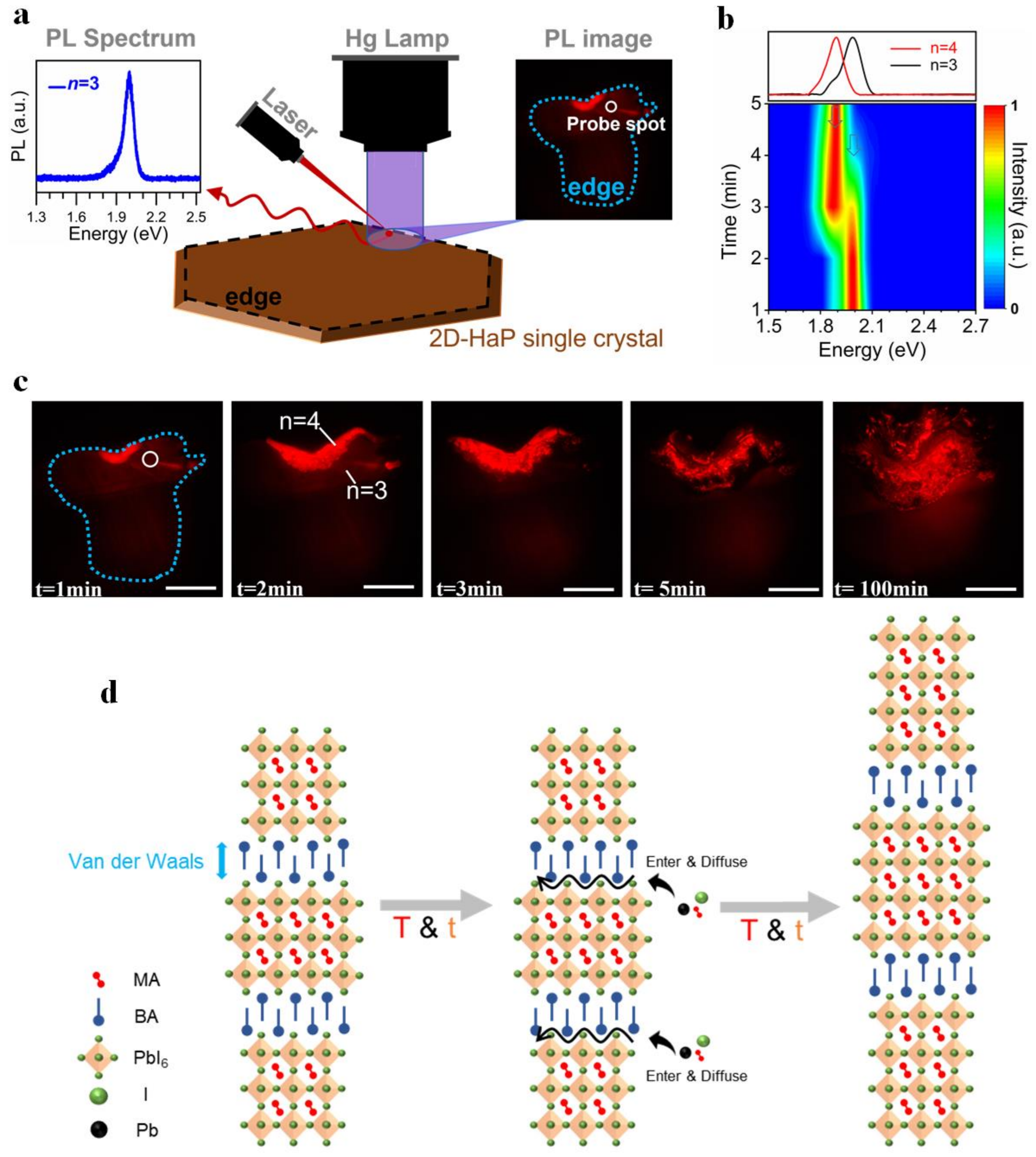

**Fig. 2 | Mechanism of transformation of 2D RPPs (*n*=3) in TCSC. (a)** In-situ PL imaging and spectrum set-up illustration, **(b)** PL spectrum intensity evolution probed at white circled spots showed in fig (a), it showed the transformation from n=3 to n=4 **(c)** In-situ Time dependent PL imaging with long pass filter (cut-off wavelength=594 nm), the crystal shape is depicted by the blue outline in the first image. PL spectrum was probed on the white circled spots. Scale bar is 100 μm. **(d)** Schematic of proposed intercalation mechanism, where Pb and MA and I diffuse through the edge and attached to the existing $[PbI_6]^{4-}$ to form new layer of $[PbI_6]^{4-}$, resulting in the increase of thickness of the inorganic

therefore quenching of the PL.[32]

To rationalize the transformation in 2D HaP from thermodynamics perspective, we calculate the enthalpy of formation (Δ H) and the Gibbs free energy (Δ G) for the RP 2D-HaP crystal. The enthalpy of transformation is calculated based on the reaction scheme Table S1, where the reactants are a specific n-value crystal, $PbI_2$, MACl while the product is the (*n*+1) number crystal (Fig. S6, Table S3-S5). Since $\Delta G = \Delta H - T(\Delta S)$, for the temperature dependent transformation, we have that $\Delta G$ is mainly dependent on the $\Delta H$ since the change of entropy $\Delta S$ is negligible[20]. The negative $\Delta H$ value of -62.52 KJ/mol for the n=3 to n=4 transformation indicates it is favorable. For the transformation of n=4 to n=5, the $\Delta H$ becomes more positive indicating a low probability of fabricating a phase pure 2D perovskite. Furthermore, the transformation from n=5 to n=6 is extremely unfavorable because of the high enthalpy of transformation, equaling to 149 KJ/mol (Table S3). To overcome this huge energy barrier, we annealed the n=5 solution at very high temperature (105 °C) for a long time (~12 hours) and managed to synthesize the large pure n=6 crystal (Fig. S7).

To further understand the relationship between the time, temperature, and phase purity and to generalize the transformation behavior, we created a phase diagram using support vector machine (SVM) classifier. A key challenge in performing machine learning analysis on perovskite materials is the availability of reproducible and insufficient data.[33] To alleviate this issue, we designed a high throughput experiment which entails 4 steps: (i) synthesizing the crystal using same parent solution but with varying temperature and time; (ii) performing large area absorption measurement on the crystal from step (i) on 3 different sample, (iii) analyzing the excitonic absorption peak position to determine the n-value; (iv) deploying ML analysis to map the phase diagram of n-value purity and different conditions of time and temperature. In total, we performed 250 KCSC synthesis for both RP and DJ. The phase diagram of RP phase and DJ phase are illustrated in fig. 3b and fig. 3c, respectively. The parent solution is of n=3 stoichiometry for both types, and the resulting crystal is classified into 4 categories: n=mixed, n=3, n=4, and n=5. The n-mixed is defined if the crystal contains more than one *n* value indicated by the multiple excitonic absorption peaks. Our best-supervised machine learning model was a support vector machine classifier with a gaussian kernel. This classifier allows for smooth nonlinear boundaries which enables better generalization of the analysis (lower variance).For RP type, below 60 °C the crystals are phase pure n=3 irrespective of time. As we slightly increase temperature (~65 °C) the synthesized crystal at short annealing times is still phase pure but at longer annealing times the crystal becomes a mixture of phases (n=3 and n=4). This indicates a temperature higher than 65 °C is needed to trigger the intercalation process. At 75 °C, the phase diagram shows that at all times the KCSC method produces mixture of phases. From 78 to 92 °C, the KCSC method produced phase pure *n* =4 RP 2D-HaP right from the start of the annealing, which indicates that higher temperature will facilitate the formation of crystal with n-value higher than parent solution, which is consistent with our thermodynamics calculations. As we increase the annealing time pure n=4 crystal also transforms to n-mixed crystal because of intercalation. The pure n=5 crystal was acquired from few conditions at higher temperature (~95 °C) but relatively short time window. In summary, the phase diagram indicates that the temperature controls the initial n-value and time controls intercalation process. Next, we investigate the DJ 2D-HaP using the same approach which also shows a transformation behavior. As shown in the fig. 3c, a ML assisted mapping of crystals yielded by KCSC method using 3-(aminomethyl)piperidinium (3AMP) DJ n=3 solution is plotted. By tuning the temperature and time of crystallization, same parent n=3 solution can form a large range of crystals, deterministically from n=3 to n=5 with ultra-high purity. The mapping of DJ 2D-HaP shows a similar trend compared to RP 2D-HaP, where n-value increases as temperature and time of crystallization increases, indicating the same intercalation mechanism showed in figure 2, also consistent with the observation of intercalation in DJ 2D-HaP from previous reports[34,35]. The chances of synthesizing purity crystal decrease as n rises, which is demonstrated by the area decrease of ascendant n in the map. Nonetheless, one difference between the RP and DJ phase diagram is: For DJ perovskite the formation of higher n-value (n=4 and n=5) requires higher temperature and longer time. This is attributed to the smaller interlayer spacing of DJ and additional interlayer interaction, limiting the precursor diffusion.[23,36] In summary, the transformation in both RP and DJ 2D-HaP series indicates it is universally applicable to all the 2D-HaPs.

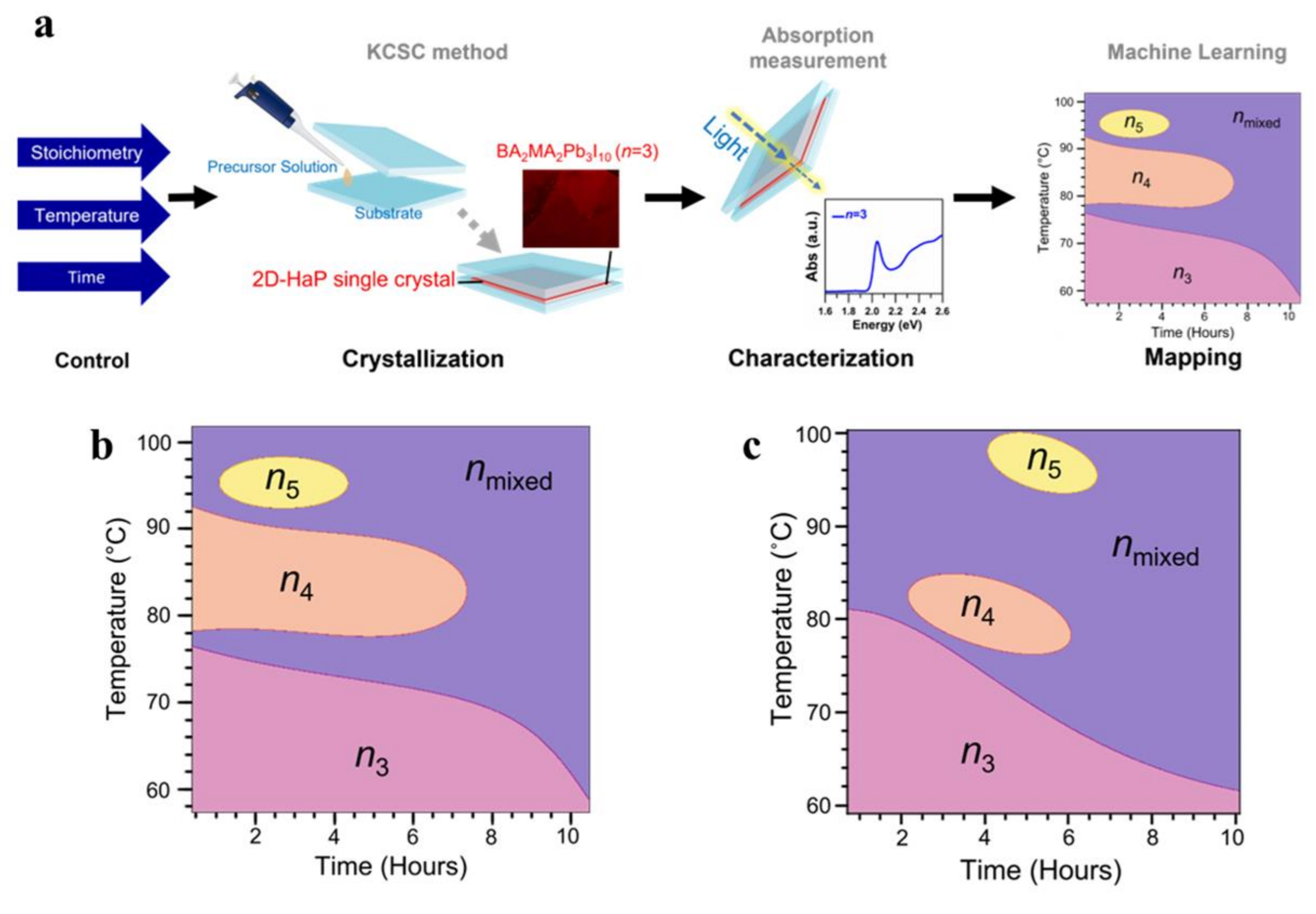

**Fig. 3 | Machine learning of transformation of 2D RPPs (*n*=3) in KCSC. (a)** ML assisted phase diagram experiment illustration. First crystal is synthesized under different temperature and time conditions, then the n-value of crystal is determined by absorbance measurement. The data was analyzed by ML to create the phase diagram. Phase diagram of RP **(b)** and DJ **(c)** n-value under different conditions of temperature and time.

To explore if this transformation principle could be scaled up, we investigated the effects of controlling the temperature and the time of classical synthesis to produce large batches of powders. Figure 4a illustrates the CS method process, in which all the precursors are mixed in a concentrated hydriodic acid (HI) at a specific stoichiometric ratio according to the target *n*-value. The solution is heated to boiling and rapidly cooled to room temperature within 1 hour or even shorter period. Small flakes with a typical lateral size of ~100μm were obtained. However, powder crystals produced from this stoichiometry-tuning CS method often contain mixtures of different n-values. For example, Fig. S8a shows the X-ray diffraction pattern for 10 classically synthesized n=3 RP powder, in which only 7 of the 10 syntheses achieved were phase-pure. The diffraction patterns of the unsuccessful synthesis show a consistent mixture of n=3 and n=4 phase (RP).

Then we apply our knowledge from transformation on the CS method, the first parameter to tune is the temperature at which the boiling solution is left to crystallize. Instead of room temperature, we test a series of higher temperatures up to 100 °C. Fig. 4b shows the powder X-ray diffraction of 2D-HaP powder crystals synthesized from 25 °C to 100 °C, which show a similar temperature dependent trend as KCSC method. Below 60 °C the n=3 will form but when the temperature rises, the resulting powder become n=3 and n=4 mixture at around 80 °C and finally pure n=4 at 100°C. However, as we showed in the fig. S8, compared to room temperature, crystallization at 55 °C has higher probability to obtain pure crystal, which is consistent with pevious discussion in which fast precipitation of the 2D perovskite crystals results in mixed phases[37] This result indicate that a temperature window of 40~55 ◦C for making reproducible phase pure 2D-HaP is critical because in this temperature range we slow crystallization rate without triggering transformation brought by higher temperature. These results again demonstrate that the temperature at which the solution is cooled to is as critical as stoichiometry in classic synthesis method. The next parameter we tune is the total amount of time in which the solution was left for crystallization. Figure 4e illustrates the X-ray diffraction patterns for various crystallization time with the temperature set to 90 ◦C. The results show a similar trend to the KCSC in which as the time increases the crystal transitions from n=3 into a mixture of n=3&n=4 and finally phase pure n=4 after 4 hours. Fig. 4d and 4f are the intensity changes of the corresponding temperature and time-dependent XRD evolution, both of them show the analogue trend where n=3 completely converts to n=4. These temperature and time dependent trends clearly indicate that stoichiometry is not the only parameter that determines the n-value of the powder 2D-HaP. The acute control of all three parameters is essential in fabricating a single n-value 2D-HaP, which we term as kinetics controlled classic synthesis (KCCS).

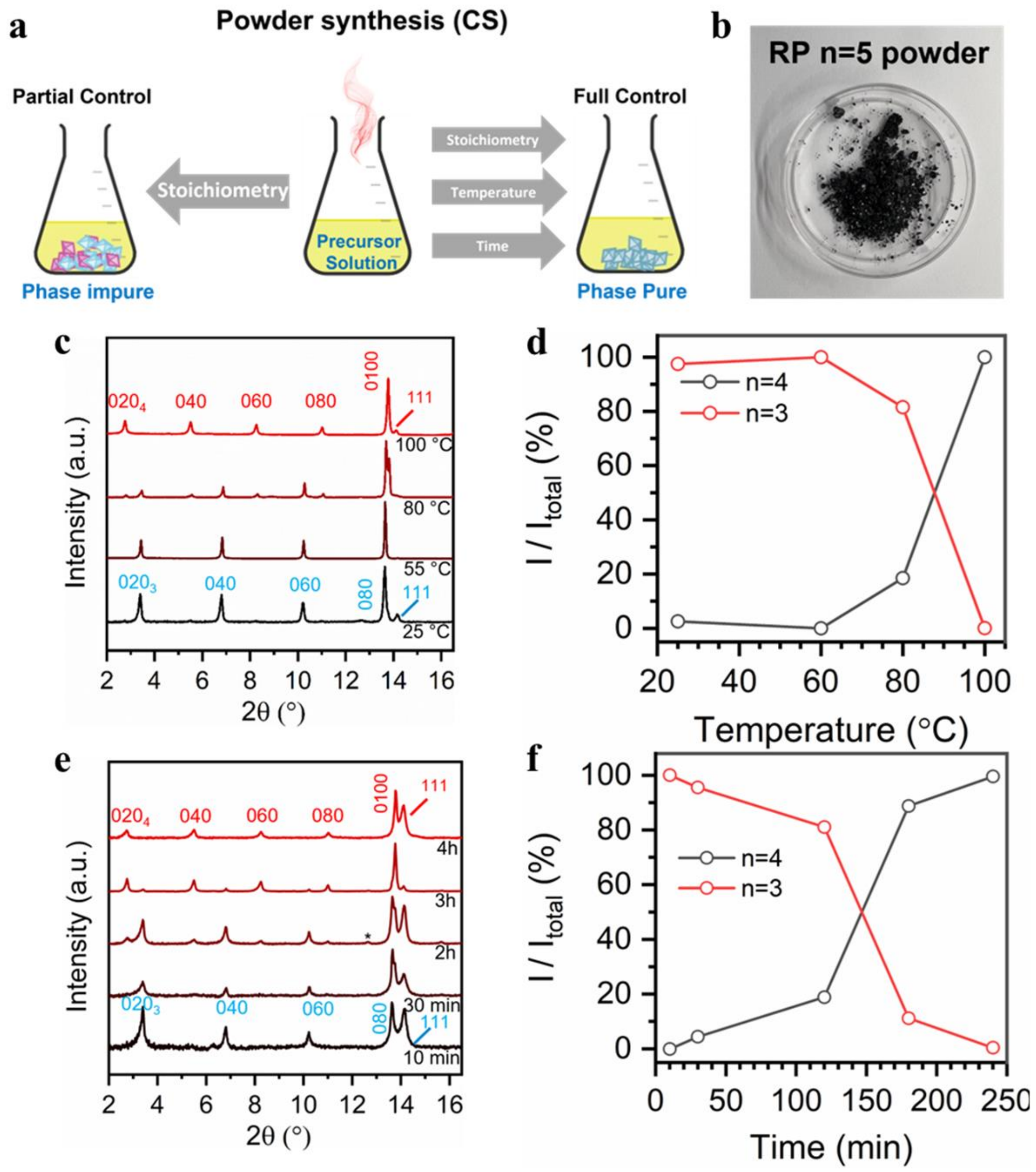

**Fig. 4| Inorganic layer thickness (n) of synthesis of 2D perovskites is temperature and rate dependent.** (**a**) Schematic illustration of the relation of classical synthesis purity and three factors determining it, which is stoichiometry, rate (time) and temperature. (**b**) Powder crystal of RP n=5; Temperature dependent (**c**) and (**d**) peak intensity evolution of CS method, it showed n increases from n=3 to n=4 as crystallization temperature rises. (**e**) time dependent 1D XRD measurement and peak intensity evolution (**f**) of CS method, it showed n=3 increase to n=4 as crystallization increases.

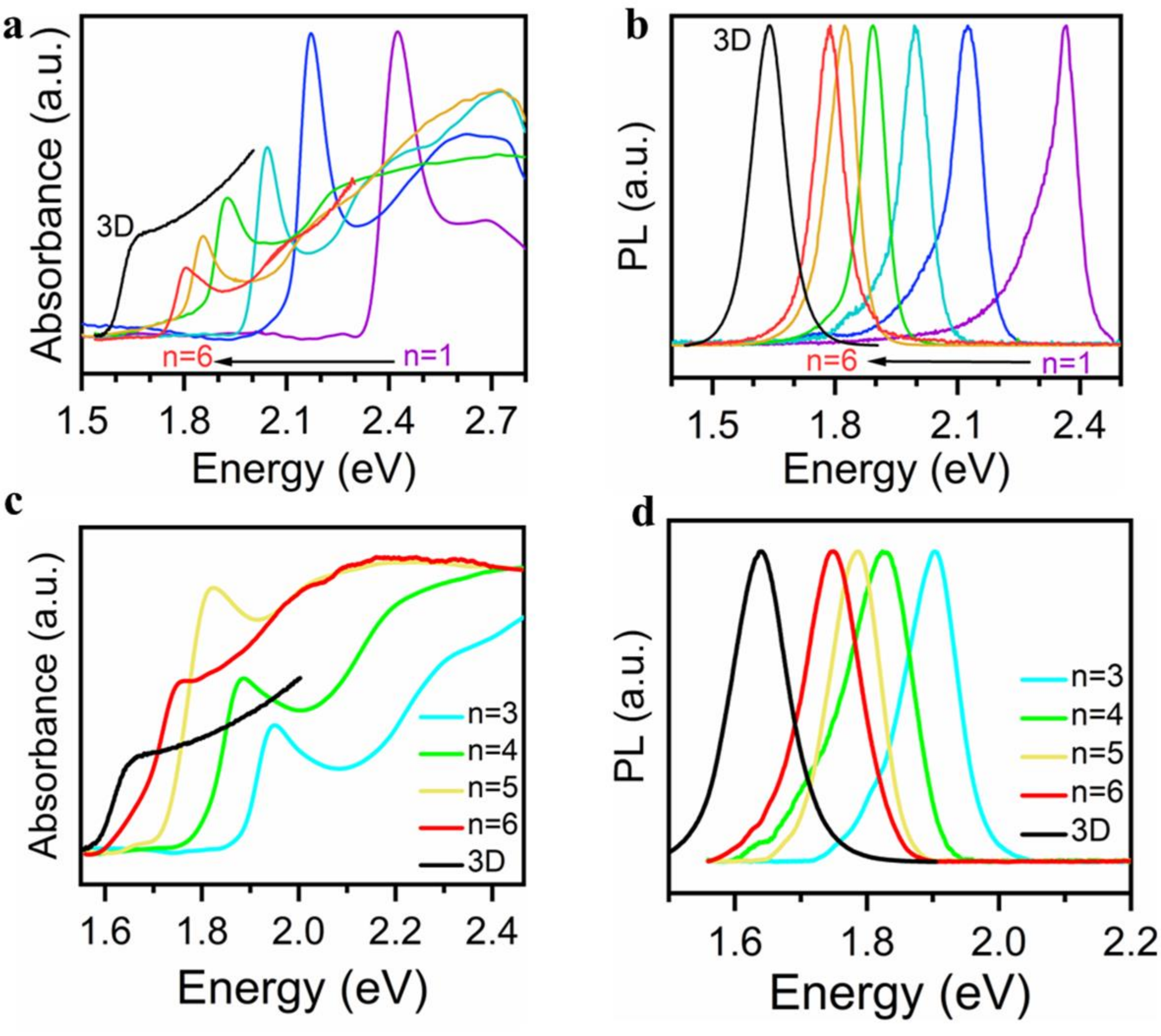

**Fig. 5 | Synthesis capability of transformation based KCSC. (a)** Absorbance, **(b)** PL of RP n=1 to n=6 and 3D MAPBI3. **(c)** Absorbance, **(d)** PL of DJ n=3 to n=6 and 3D MAPBI3.

Utilizing the knowledge of progressive transformation and the ML model, we are able to grow phase pure single crystals ranging from *n*=1 to *n*=6 for both RP DJ 2D-HaP. Figure 5a, b shows the absorption, photoluminescence spectrum of KCSC produced *n*=1 to *n*=6 RP 2D-HaP single crystal. Each of the KCSC fabricated crystals exhibit a single narrow absorption peak decreasing from 2.42eV to 1.78eV as a function of *n* value (*n*=6 to *n*=1). The photoluminescence emission peak also follows this trend from 2.38eV to 1.77eV. This behavior is consistent with their quantum and dielectric confinement effects and furthermore indicates the phase purity of the fabricated crystal[38]. Both absorption and PL indicate the ultra-high purity of the crystal synthesized by our method. The absorption and photoluminescence of a control $MAPbI_3$ 3D-HaP is shown in the black curve on both plots. 1D X-Ray diffraction of KCSC fabricated crystal (Fig. S9) also shows ultra-high purity. Figure 5c, d shows the absorption, photoluminescence spectrum of KCSC produced *n*=3 to *n*=6 DJ 2D-HaP single crystal. Absorption (fig.5c) shows the 3AMP follow the general trend for 2D-HaP where the band gap decreases as *n* increases (1.95eV for n=3, 1.88eV for n=4, 1.82eV for n=5, 1.75eV for n=6). Steady-state photoluminescence (PL) shows an identical trend with the band gaps (1.90eV for n=3, 1.82eV for n=4, 1.79eV for n=5, 1.75eV for n=6). Similar to the 2D-HaP synthesized from KCSC and transformation, the purity of the DJ 2D-HaP synthesized by same approach is of ultra-high purity. The band gap of DJ 3AMP series is smaller than RP BA series which is consistent with previous reports. [2,36,39] Transformation of the 4-(aminomethyl)piperidinium (4AMP) 2D-HaP is shown in Fig. S10.

**Acknowledgments:** The work at Rice University was supported by start-up funds under the molecular nanotechnology initiative and also the DOE-EERE 2022-1652 program. J.H. acknowledges the financial support from the China Scholarships Council (No. 202107990007). W.L. acknowledges the National Science Foundation Graduate Research Fellowship Program. J.E. acknowledges the financial support from the Institut Universitaire de France.

**Author contributions:** J.H., J.-C.B. and A.D.M. conceived and designed the experiment. J.H. synthesized the perovskite single crystals with the help of S.S.. J.H. and H.Z. performed optical characterizations with the help of W.L.. J.H. performed the transformation experiment. W.L. performed the machine learning analysis. J.H. performed data analysis with guidance from J.E., J.-C.B., and A. D. M.. J.H. and W.L. wrote the manuscript with input from everyone. All authors read the manuscript and agree to its contents, and all data are reported in the main text and supplemental materials.

**Competing interests:** The authors declare no competing interests.

Supplementary Information for

# Deterministic synthesis of phase pure 2D perovskites via progressive transformation of layer thickness

Jin Hou[1], Wenbin Li[2,3], Hao Zhang[2,3], Siraj Sidhik[1], Jean-Christophe Blancon[2], Jacky Even[4], Mercouri G. Kanatzidis[5] and Aditya D. Mohite[1,2*]

**[1]Department of Materials Science and NanoEngineering, Rice University, Houston, Texas 77005, USA.**

**[2]Department of Chemical and Biomolecular Engineering, Rice University, Houston, Texas 77005, USA.**

**[3]Applied Physics Graduate Program, Smalley-Curl Institute, Rice University, Houston, TX, 77005, USA.**

**[4]Fonctions Optiques pour les Technologies de l'Information (FOTON), INSA de Rennes, CNRS, UMR 6082, 35708 Rennes, France.**

**[5]Department of Chemistry and Department of Materials Science and Engineering, Northwestern University, Evanston, Illinois 60208, USA.**

## Supplementary Figures

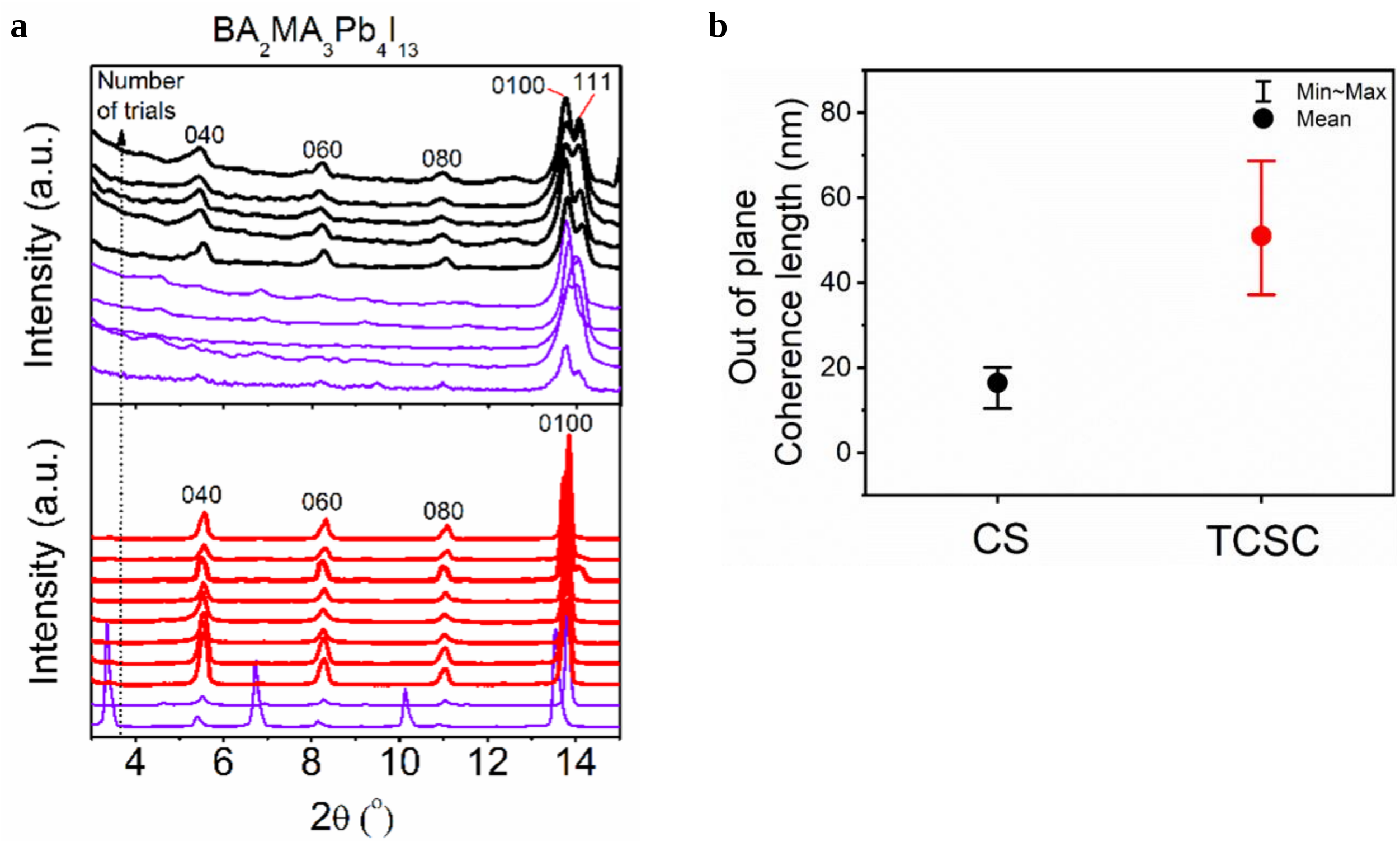

**Fig. S1. (a)** Comparison of Reproducibility by 1D X-Ray diffraction measurement of $BA_2MA_3Pb_4I_{13}$ crystals synthesized by both methods. (**b)** Out-of-plane coherence length (estimated grain size) extracted from the full-width-at-half-max of 1D X-ray diffraction pattern of the (040) peaks of both CS and KCSC methods.

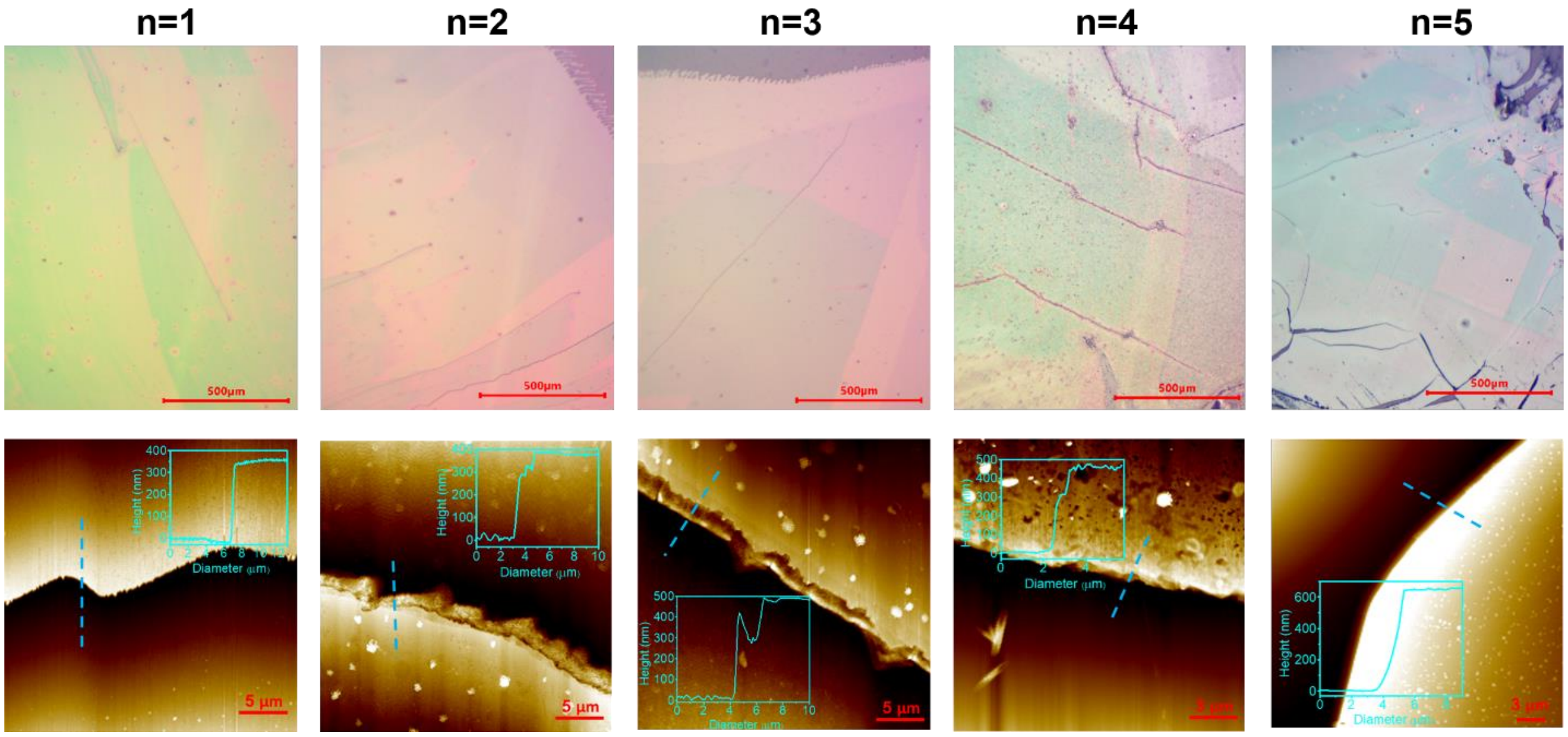

**Supplementary fig. 1.** Microscope images and AFM images of KCSC RP n=1 to 5 crystals.

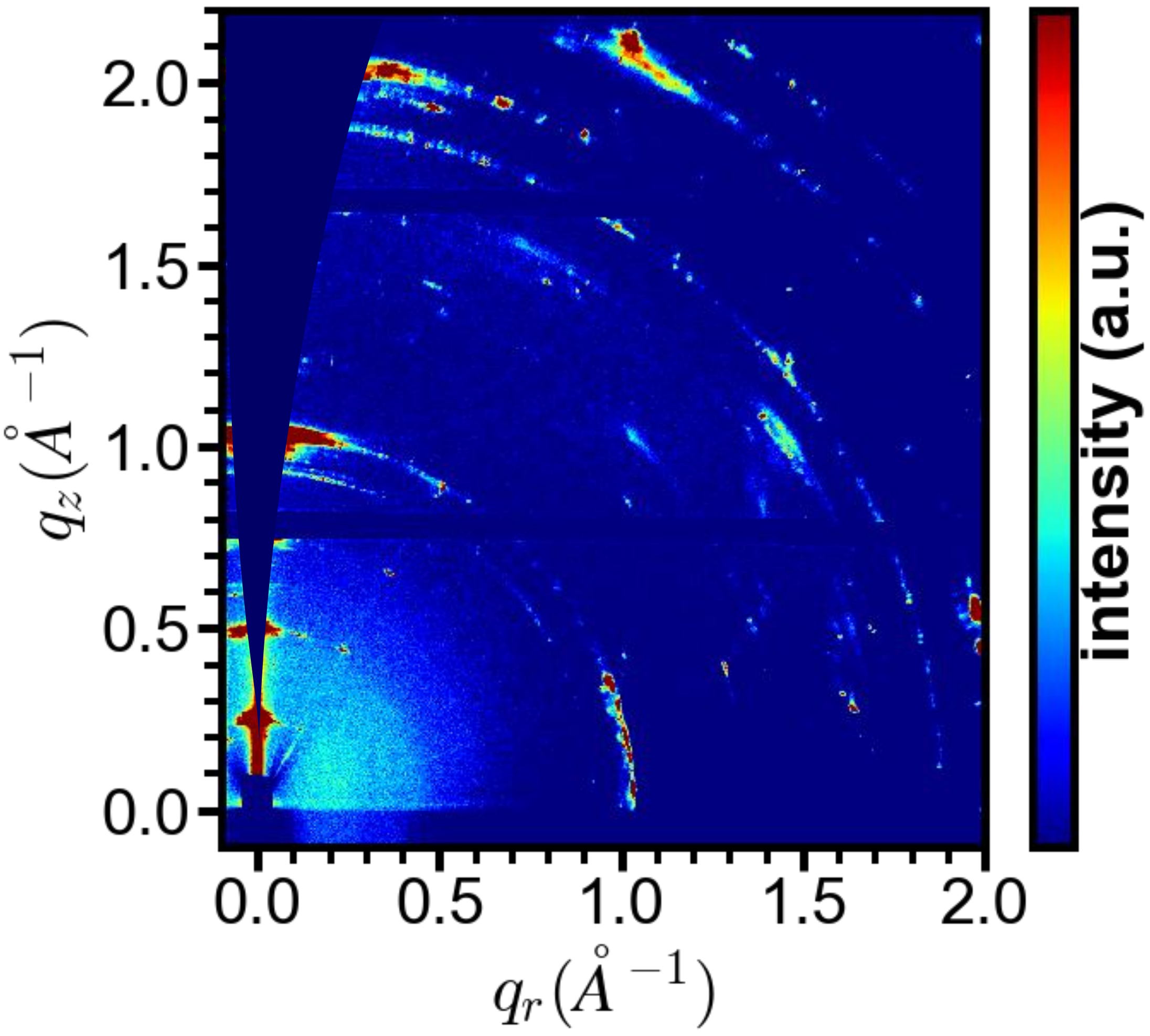

**Supplementary fig. 2: Grazing incident wide angle X-ray scattering pattern of the RP n=3 TCSC sample.** The spot-like diffraction pattern indicates a single crystalline sample with large grain boundaries, the diffraction spots on the qz direction illustrates a horizontally oriented perovskite layer (with respect to the substrate).

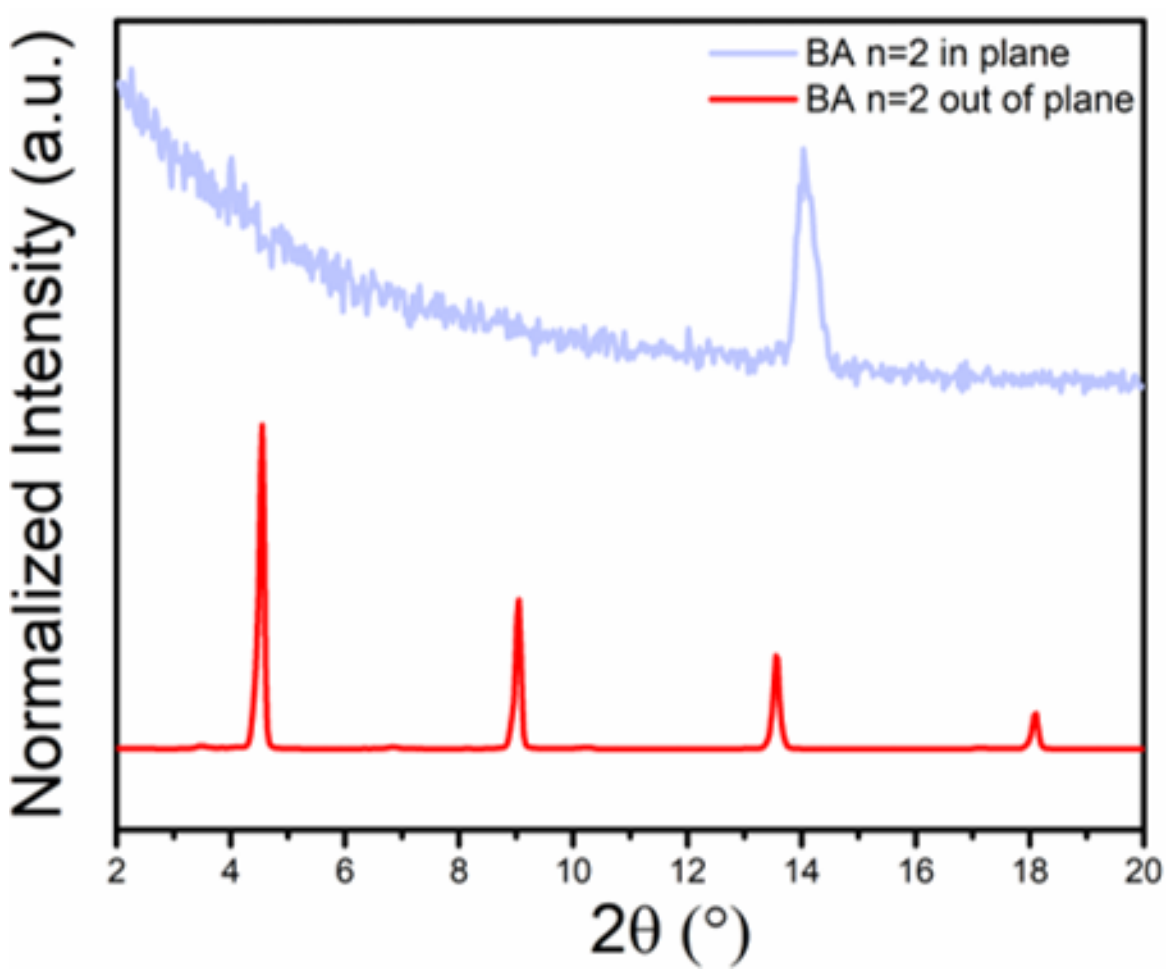

**Supplementary fig. 3.** The In-plane and out of plane 1D XRD of RP *n*=2 KCSC indicated the perfect horizontal orientation of KCSC crystal. The In-Plane XRD only shows the (0k0) peaks while the Out-of-Plane XRD only shows the (111) peak.

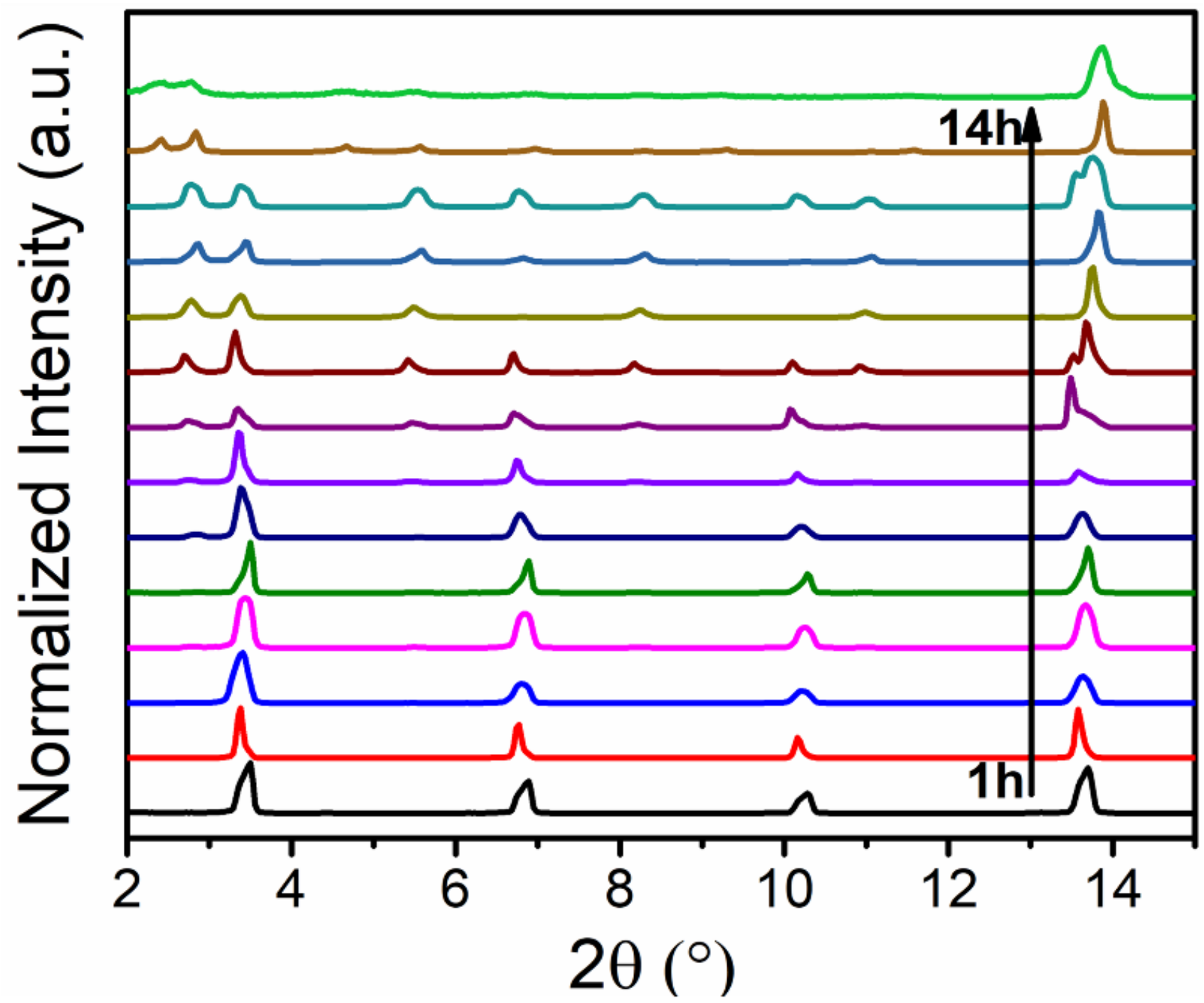

**Supplementary fig. 4:** n=3 parent solution time dependent transformation at 75 °C.

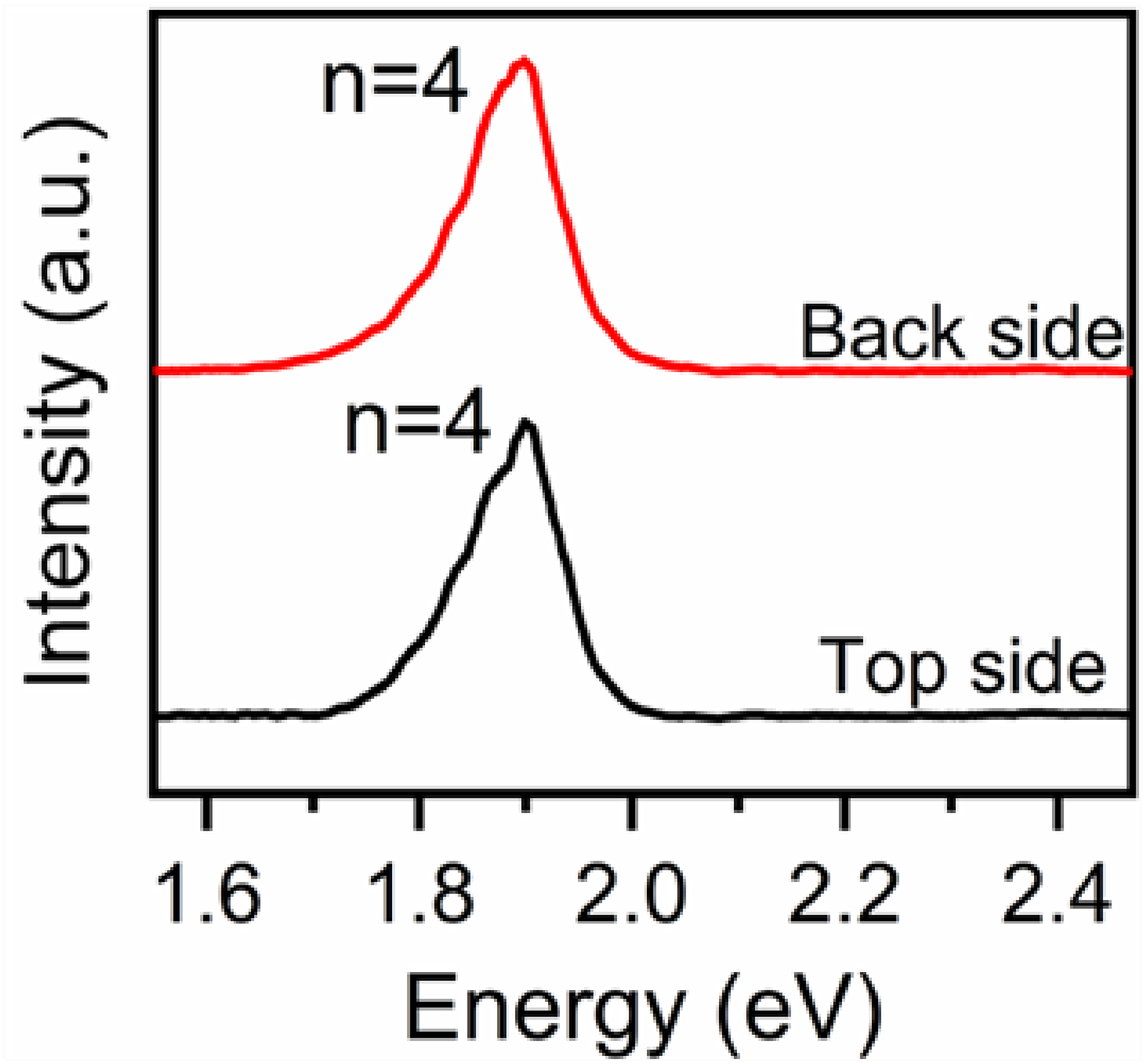

**Supplementary fig. 5:** Back side PL spectrum of a spot where the top-side PL showed complete transformation to n=4, it showed identical transformation to n=4 as well.

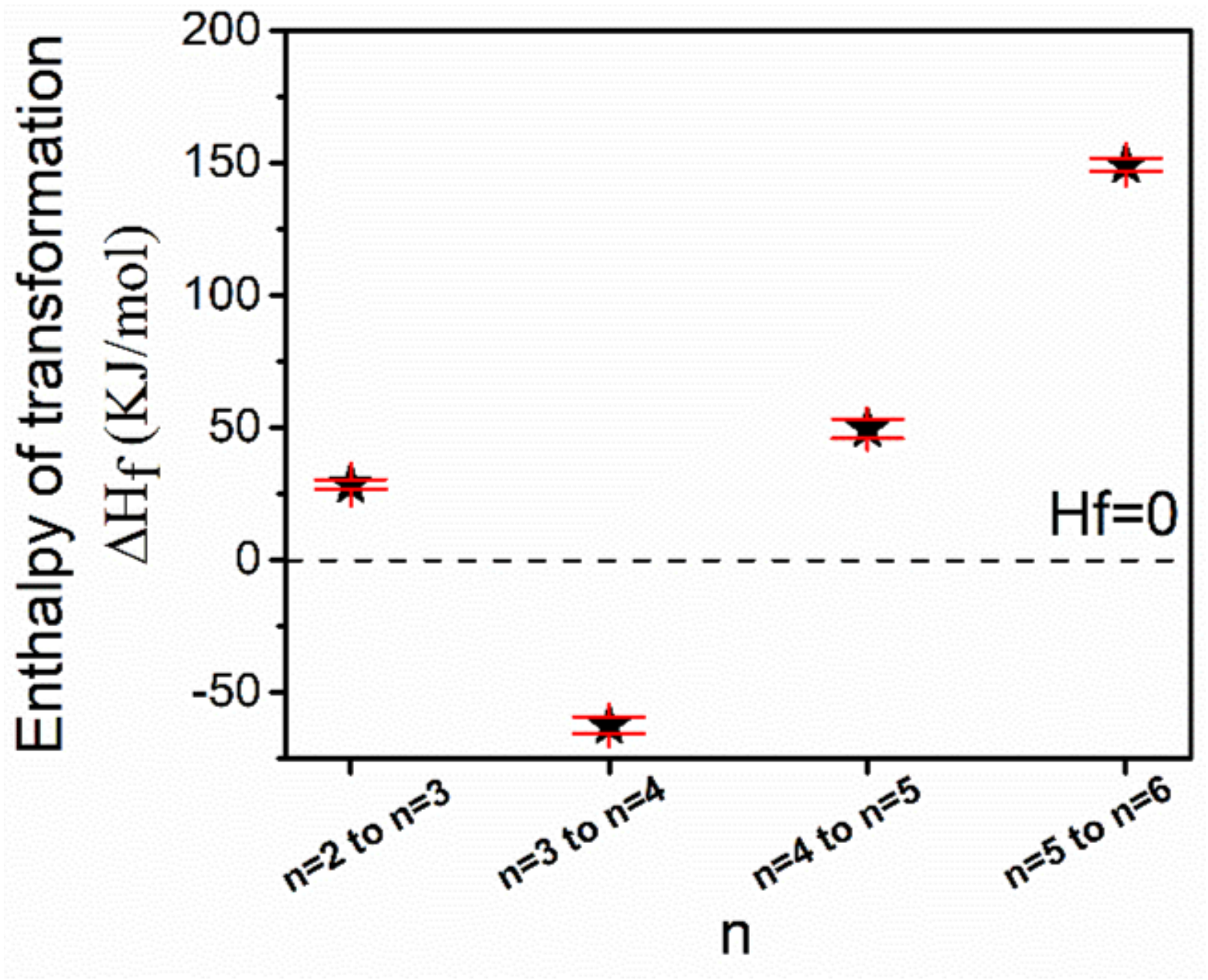

**Supplementary fig. 6: (a)** Transformation enthalpy of each n of RPPs synthesized by its lower n to the higher.

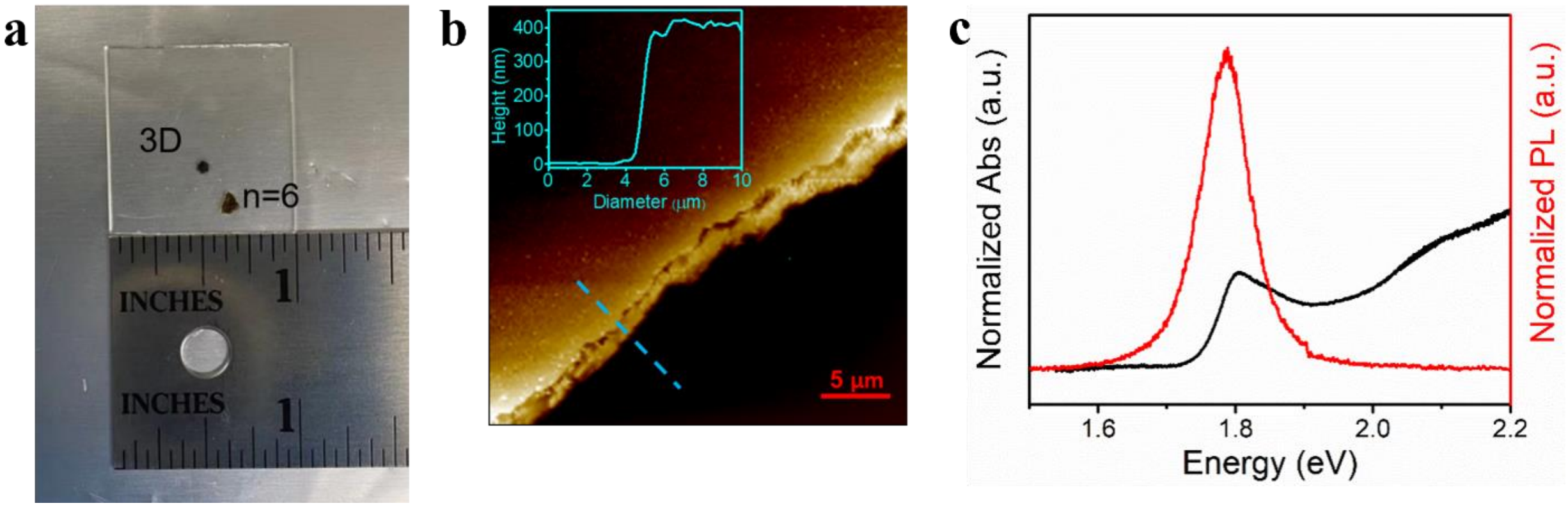

**Supplementary fig 7. (a)** The photo, **(b)** AFM image **(c)** Normalized absorption and Photoluminescence intensity of n=6 RPPs millimeter sized crystal produced from TCSC method using n=5 parent solution.

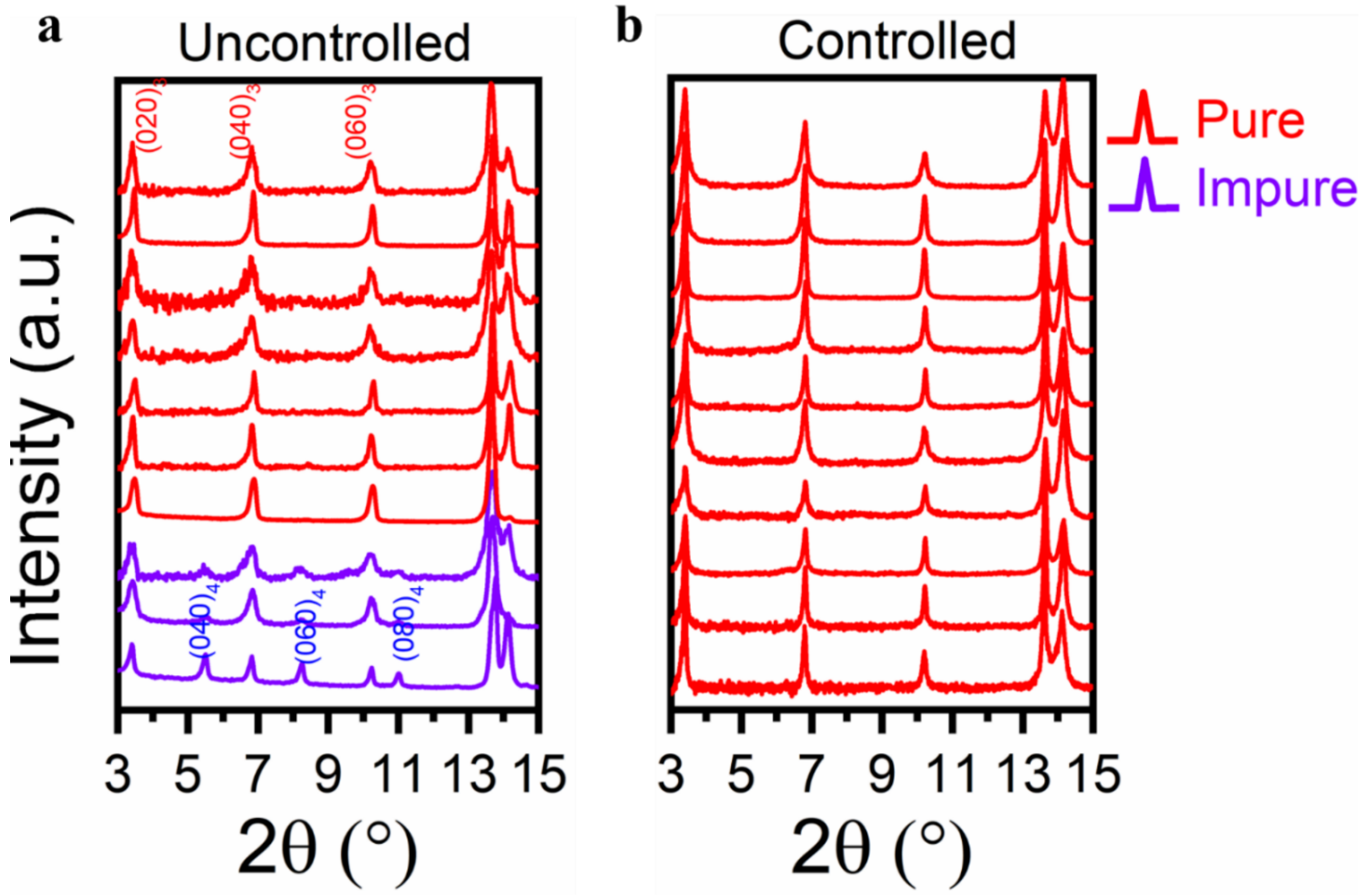

**Supplementary fig. 8: (a)** Uncontrolled CS showed 70% percent of success rate when synthesizing RPP n=3 when the (**b**) controlled the CS showed a 100% success rate.

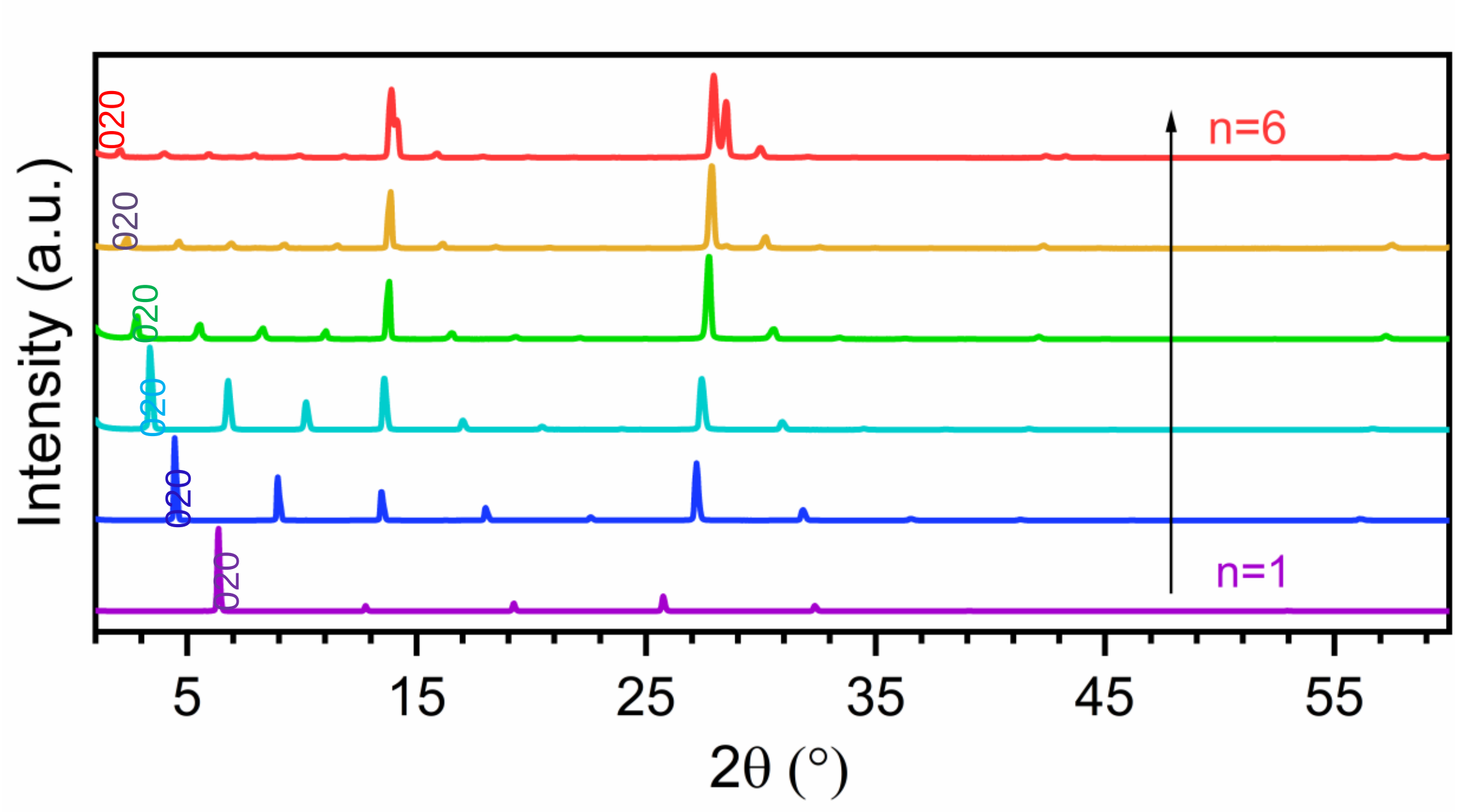

**Supplementary fig. 9:** 1D X-Ray diffraction pattern of of RPPs single crystal (n=1 to 6) produced by KCSC.

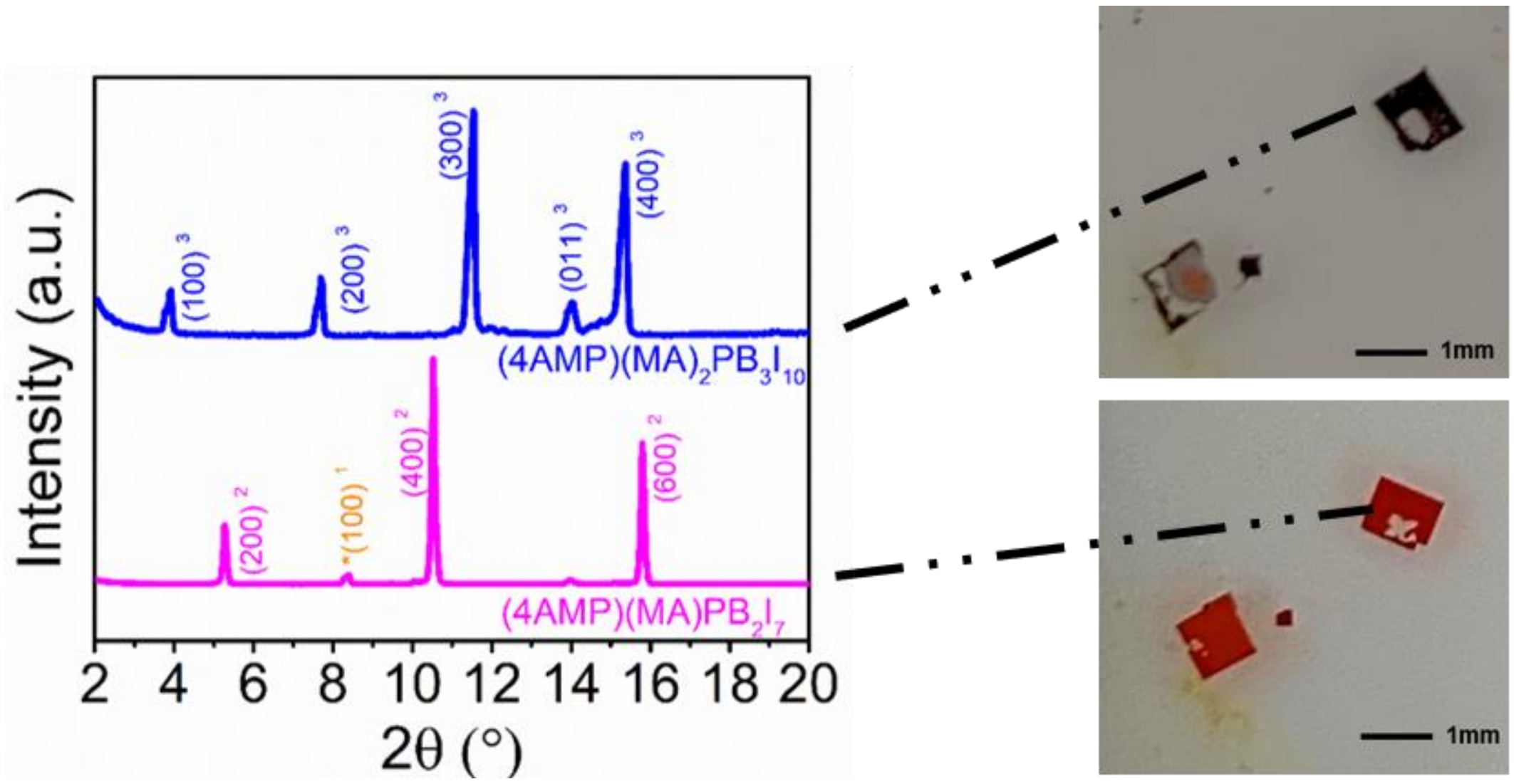

**Supplementary fig. 10: Dion-Jacbson 4AMP perovskites transformation from n=2 to n=3.** The XRD peaks evolutions and the clear color change from red to black indicating the DJ 4amp n=2 perovskites transformed to n=3 by using KCSC method.

## Supplementary tables

**Table S1.** Thermochemical cycle used to calculate enthalpy of transformation RP (n) to RP (n+1).

| | |
|---|---|
| $(BA)_2(MA)_{n-1}Pb_nI_{3n+1(S)}$ (n) $\rightarrow$ $PbI_{2(aq)}$ + $2BAI_{(aq)}$ + (n-1)$MAI_{(aq)}$ | $\Delta H_1$ |
| $(BA)_2(MA)_{n+1-1}Pb_{n+1}I_{3(n+1)+1(S)}$ (n+1) $\rightarrow$ (n+1)$PbI_{2(aq)}$ + $2BAI_{(aq)}$ + (n)$MAI_{(aq)}$ | $\Delta H_2$ |
| $PbI_{2(S)} \rightarrow PbI_{2(aq)}$ | $\Delta H_3$ |
| $MAI_{(S)} \rightarrow MAI_{(aq)}$ | $\Delta H_4$ |
| $(BA)_2(MA)_{n-1}Pb_nI_{3n+1(S)}$ (n) + $PbI_{2(S)}$ +$MAI_{(S)}$ $\rightarrow$ $(BA)_2(MA)_{n+1-1}Pb_{n+1}I_{3(n+1)+1(S)}$ (n+1) | $\Delta H_f$ |

$\Delta H_f = \Delta H_1 - \Delta H_2 + \Delta H_3 + \Delta H_4$

**Table S2.** Thermochemical data used to calculate enthalpy of transformation RP (n) to RP (n+1)

| Compound | n | Hs (KJ/mol) |
|---|---|---|
| $PbI_2$ | - | 33.28±0.55 (3) |
| MACl | - | 8.96±0.15 (4) |
| $BA_2MAPb_2I_7$ | 2 | 160.74±1.36 (3) |
| $BA_2MA_2Pb_3I_{10}$ | 3 | 174.46±1.12 (3) |
| $BA_2MA_3Pb_4I_{13}$ | 4 | 279.22±3.06 (3) |
| $BA_2MA_4Pb_5I_{16}$ | 5 | 271.95±2.02 (3) |
| $BA_2MA_5Pb_6I_{19}$ | 6 | 164.77±1.34 (3) |

**Table S3.** Calculated enthalpy of transformation RP (n) to RP (n+1)

| Transformation | Enthalpy of transformation (KJ/mol) |
|---|---|
| RPPs n=2 to n=3 | 28.52±1.85 |
| RPPs n=3 to n=4 | -62.52±3.31 |
| RPPs n=4 to n=5 | 49.51±3.71 |
| RPPs n=5 to n=6 | 149.42±2.49 |

**Table S4.** Propagation of uncertainty mathematical functions used for calculation

| Function | $U_R$ |
|---|---|
| R=kA | $u_R=ku_A$ |
| R=A+B | $u_R= \sqrt{u_A^2 + \mu_B^2}$ |
| R=A-B | $u_R= \sqrt{u_A^2 + \mu_B^2}$ |

**Table S5.** The overall performance of the support vector machine (SVM)

| Classes | Precision | Recall | F1-score |
|---|---|---|---|
| **Mixed purity** | 0.82 | 0.74 | 0.78 |
| **n=3** | 0.94 | 0.81 | 0.87 |
| **n=4** | 0.72 | 0.93 | 0.81 |
| **n=5** | 0.71 | 1 | 0.82 |
| | | | |
| **Average Percision** | | 0.828028169 | |
| **Average Recall** | | 0.816478873 | |